\documentclass[prb,preprint]{revtex4-1} 

\usepackage{amsfonts} 
\usepackage{graphicx} 
\usepackage{amsmath} 
\usepackage{amssymb} 
\usepackage{amsthm} 
\usepackage{bm} 
\usepackage{color} 
\usepackage{amstext}
\usepackage{hyperref}

\renewcommand{\Re}{\operatorname{Re}}
\renewcommand{\Im}{\operatorname{Im}}

\begin{document}

\title{A Completely Algebraic Solution of the Simple Harmonic Oscillator}

\author{M.~Rushka}
\email{mjr294@georgetown.edu} 
\author{J.~K.~Freericks} 
\email{james.freericks@georgetown.edu}
\affiliation{Department of Physics, Georgetown University, 37th and O Sts. NW, Washington, DC 20057 USA}

\date{\today}

\begin{abstract}
We present a full algebraic derivation of the wavefunctions of the simple harmonic oscillator in coordinate and momentum space. This derivation illustrates the abstract approach to the simple harmonic oscillator by completing the derivation of the representation-dependent wavefunctions from the representation-independent energy eigenfunctions. It is simple to incorporate into the undergraduate and graduate curricula. This new derivation begins with the standard approach that was first presented by Dirac in 1947 (and is modified slightly here in the spirit of the Schr\"odinger factorization method), and then supplements it by employing the translation (or boost) operator to determine the wavefunctions algebraically, without any derivatives. In addition, we provide a summary of the history of this approach, which seems to have been neglected by most historians of quantum mechanics, until now.
\end{abstract}

\maketitle 

\section{Introduction} 

The Hamiltonian of the simple harmonic oscillator is 
\begin{equation}
    \hat{\mathcal{H}} = \frac{\hat{p}^2}{2m}+\frac{1}{2}m\omega_0^2\hat{x}^2,
\end{equation}
where $\hat{p}$ and $\hat{x}$ denote the momentum and position operators, which satisfy the canonical commutation relation 
\begin{equation}
[\hat x,\hat p]=\hat x\hat p-\hat p\hat x=i\hbar
\end{equation} 
(hats will be used on all operators throughout this work). Here, we have $m$ the mass and $\omega_0$ the frequency of the oscillator. Most textbooks solve this problem in two ways: (1) first, one represents the momentum operator via $\hat p=-i\hbar \,\frac{d}{dx}$ and solves the resulting differential equation, finding the energy eigenvalues via the condition that the solution be bounded as $|x|\to\infty$; and (2) an abstract operator method is employed to factorize the Hamiltonian and then used to determine the energies and an operator form of the eigenfunctions of the different energy eigenstates. When it comes time to determine the wavefunctions in the latter case, one converts the lowering operator into the coordinate-space representation, which yields a first-order differential equation for the ground state. Then applying the raising operators in the coordinate representation to the ground state produces the excited state wavefunctions in coordinate space; a similar approach can also be used in momentum space. While this approach is time-tested and familiar, it seems unbalanced to require going to the coordinate representation to calculate the wavefunctions. Indeed, this is not necessary, and we show that an algebraic derivation of the wavefunctions is possible. We believe that this should become part of standard treatment of the simple harmonic oscillator. We do want to point out that the textbook by Bohm does hint at alternatives to the ``standard'' approach\cite{bohm93} but does not proceed in the fashion we propose here. As we will see, the key to our procedure lies in the proper use of translation operators, which are often discussed in quantum textbooks, but hardly used for anything meaningful (except sometimes to derive the canonical commutation relation between $\hat x$ and $\hat p$).

\section{Schr\"odinger Factorization Method}

Before jumping into the derivation, we briefly summarize the Schr\"odinger factorization method following the textbooks of Green\cite{Green65} and Ohanian\cite{ohanian} because the method is not well known to many (Schr\"odinger's original reference is also quite readable\cite{schroedinger4041}). We do so here to present the context for our slight change in the standard algebraic derivation of the simple harmonic oscillator eigenstates. We employ the Dirac notation for states in the Hilbert space throughout this work.

While Schr\"odinger's discovery of the Schr\"odinger equation is widely known today, his work from the 1940s on the so-called factorization method is less familiar. This portion of Schr\"odinger's work has been omitted from most quantum textbooks with the exception of its application to the harmonic oscillator, the simplest example of this technique. The general factorization method may appear rather abstract, but it can be straightforwardly applied to an array of problems. In fact, any problem that can be solved via Schr\"odinger's differential equation can also be solved using the factorization method.

The method works by factorizing the Hamiltonian $\hat{\mathcal{H}}$. However, unlike the operator approach to the harmonic oscillator, the factorization method then constructs a set of auxiliary Hamiltonians $\hat{\mathcal{H}}_j$ which differ from the initial Hamiltonian. These auxiliary Hamiltonians and their subsequent factorizations are used to determine the energy eigenstates of the original Hamiltonian. We begin by finding a set of lowering and raising operators $\hat{A}_j$ and $\hat{A}_j^{\dagger}$ (with $j=0, \, 1...$), which factorize the auxiliary Hamiltonians as follows:
\begin{eqnarray}
\hat{\mathcal{H}} = \hat{\mathcal{H}}_0 &=& \hat{A}_0^{\dagger}\hat{A}_0^{\phantom\dagger} + E_0 \\
\hat{\mathcal{H}}_1 &=& \hat{A}_1^{\dagger}\hat{A}_1^{\phantom\dagger} + E_1 \\
&\vdots& \nonumber \\
\hat{\mathcal{H}}_j &=& \hat{A}_j^{\dagger}\hat{A}_j^{\phantom\dagger} + E_j \label{eq: auxiliaryH}
\end{eqnarray}
where the energies $E_j$ are scalars; we suppress identity operators multiplying the numbers $E_j$ throughout. Note that we use capital letters to denote Schr\"odinger raising and lowering operators, as opposed to the Dirac versions which are lower case and will be introduced below. We determine the auxiliary Hamitonians $\hat{\mathcal{H}}_j$ via the additional defining relations
\begin{equation}
    \hat{\mathcal{H}}_j = \hat{A}_{j-1}^{\phantom\dagger}\hat{A}_{j-1}^{\dagger} + E_{j-1}. \label{eq: SFM_1}
\end{equation}
Equation~(\ref{eq: SFM_1}) ensures that we can construct all the auxiliary Hamiltonians starting from $\hat{\mathcal{H}}_0, \, \hat{A}_0^{\dagger}, \, \hat{A}_0^{\phantom\dagger},$ and $E_0$. We make this procedure well-defined by choosing the largest possible value for the energies if we find any ambiguity in selecting the $E_j$'s. This requirement prevents us from setting both $\hat{A}_j^{\phantom\dagger} = \hat{A}_{j-1}^{\dagger}$ and $E_j=E_{j-1}$, and hence the $E_j$ form a nondecreasing sequence
\begin{equation}
    E_0 \leq E_1 \leq ... \leq E_{j-1} \leq E_j.
\end{equation}
Note that the procedure for choosing the maximal $E_j$ is simply stated in Bert Green's book~\cite{Green65} without proof---we are unable to establish why this is the correct criterion, but it does work for all known examples. We anticipate that it arises from the condition that the auxiliary Hamiltonian have a normalizable ground state.

We first establish that $E_j$ is an eigenvalue of the original Hamiltonian $\hat{\mathcal{H}}$. We let $|\psi\rangle$ denote some eigenfunction of $\hat{\mathcal{H}}$ with eigenvalue $E$, so that $\hat{\mathcal{H}}|\psi\rangle = E|\psi\rangle$.
We claim that if there exists some $j$ such that if $E=E_j$, then there exists a state
\begin{equation}
    \hat{A}_j^{\phantom\dagger}\hat{A}_{j-1}^{\phantom\dagger}\hat{A}_{j-2}^{\phantom\dagger}...\hat{A}_1^{\phantom\dagger}\hat{A}_0^{\phantom\dagger} | \psi \rangle .
\end{equation}
If this is not the case, then $E$ is larger than or equal to the maximum discrete eigenvalue $E_j^{\text{max}}$. To prove this claim, we begin with the set of states $|\phi_j\rangle$, which satisfy
\begin{equation}
    |\phi_j \rangle = \hat{A}_j^{\phantom\dagger}\hat{A}_{j-1}^{\phantom\dagger}\hat{A}_{j-2}^{\phantom\dagger}...\hat{A}_1^{\phantom\dagger}\hat{A}_0^{\phantom\dagger} | \psi \rangle 
\end{equation}
for $j=0,1,2,\ldots$. Then we use the intertwining relationship for the auxiliary Hamiltonians, given by
\begin{equation}
    \hat{\mathcal{H}}_{j+1}\hat{A}_j^{\phantom\dagger} = \hat{A}_j^{\phantom\dagger}\hat{\mathcal{H}}_j. \label{eq: SFM_2}
\end{equation}
This relationship is established by using the definition of the auxiliary Hamiltonians and Eq.~(\ref{eq: SFM_1}) via
\begin{eqnarray}
\hat{\mathcal{H}}_{j+1}\hat{A}_j^{\phantom\dagger} &=& (\hat{A}_{j+1}^{\dagger}\hat{A}_{j+1}^{\phantom\dagger}+E_{j+1})\hat{A}_j^{\phantom\dagger} \nonumber\\
&=& (\hat{A}_j^{\phantom\dagger}\hat{A}_j^{\dagger}+E_j)\hat{A}_j^{\phantom\dagger} \nonumber\\
&=& \hat{A}_j^{\phantom\dagger}\hat{A}_j^{\dagger}\hat{A}_j^{\phantom\dagger}+E_j\hat{A}_j^{\phantom\dagger} \nonumber\\
&=& \hat{A}_j^{\phantom\dagger}(\hat{A}_j^{\dagger}\hat{A}_j^{\phantom\dagger}+E_j) \nonumber\\
&=& \hat{A}_j^{\phantom\dagger}\hat{\mathcal{H}_j}.
\end{eqnarray}
We now compute the norm squared
\begin{equation}
    \langle \phi_j | \phi_j \rangle = \langle \psi| \hat{A}_0^{\dagger}\hat{A}_1^{\dagger}...\hat{A}_j^{\dagger}\hat{A}_j^{\phantom\dagger}...\hat{A}_1^{\phantom\dagger}\hat{A}_0^{\phantom\dagger} | \psi \rangle
\end{equation}
which is always nonnegative. For $j=0$, we have
\begin{equation}
    \langle \phi_0 | \phi_0 \rangle = \langle \psi| \hat{A}_0^{\dagger}\hat{A}_0^{\phantom\dagger}|\psi\rangle = \langle \psi |\left ( \hat{\mathcal{H}}-E_0\right )|\psi\rangle = (E-E_0) \geq 0, \label{eq: SFM_3}
\end{equation}
which follows due to $\hat{\mathcal{H}}|\psi\rangle=E|\psi\rangle$ and $\langle\psi|\psi\rangle=1$. For $j=1$, we find
\begin{eqnarray}
\langle \phi_1|\phi_1\rangle &=& \langle \psi | \hat{A}_0^{\dagger}\hat{A}_1^{\dagger}\hat{A}_1^{\phantom\dagger}\hat{A}_0^{\phantom\dagger}|\psi|\rangle\nonumber \\
&=& \langle \psi| \hat{A}_0^{\dagger}(\hat{\mathcal{H}_1}-E_1)\hat{A}_0^{\phantom\dagger}|\psi\rangle \nonumber\\
&=& \langle \psi|(\hat{\mathcal{H}}-E_1)\hat{A}_0^{\dagger}\hat{A}_0^{\phantom\dagger}|\psi\rangle \nonumber\\
&=& \langle \psi|(\hat{\mathcal{H}}-E_1)(\hat{\mathcal{H}}-E_0)|\psi\rangle \nonumber\\
&=& (E-E_1)(E-E_0) \geq 0. \label{eq: SFM_4}
\end{eqnarray}
Here we use the intertwining relation $\hat{A}_0^{\dagger}\hat{\mathcal{H}_1} = \hat{\mathcal{H}}_0\hat{A}_0^{\dagger}$, which follows from Hermitian conjugation of Eq.~(\ref{eq: SFM_2}). We can extend these results to all $j$ to find
\begin{equation}
    \langle \phi_j | \phi_j \rangle = (E-E_j)(E-E_{j-1})...(E-E_1)(E-E_0) \geq 0. \label{eq: SFM_5}
\end{equation}
Note that from Eq.~(\ref{eq: SFM_3}) we have $E \geq E_0$ and from Eq.~(\ref{eq: SFM_4}) that either $E=E_0$ or $E \geq E_1$. Again, we can extend this argument to all $j$ and show that either $E=E_j$ for some $j$, or $E \geq E_j^{\text{max}}$, in which case $E$ is in the continuous part of the spectrum. We have thus found that the discrete eigenvalues of $\hat{\mathcal{H}}$ are the energies $E_j$.

It remains to find the eigenfunctions of $\hat{\mathcal{H}}$. For $E=E_j$, we have
\begin{equation}
    \langle \phi_j | \phi_j \rangle = \langle \phi_{j-1} | \hat{A}_j^{\dagger}\hat{A}_j^{\phantom\dagger}|\phi_{j-1}\rangle = ||\hat{A}_j^{\phantom\dagger}|\phi_{j-1}\rangle||^2 = 0,
\end{equation}
which implies 
\begin{equation}
\hat{A}_j^{\phantom\dagger}|\phi_{j-1}\rangle=0.
\end{equation}
In addition, we have
\begin{equation}
    (\hat{\mathcal{H}}_j-E_j)|\phi_{j-1}\rangle = (\hat{A}_j^{\dagger}\hat{A}_j^{\phantom\dagger}+E_j-E_j)|\phi_{j-1}\rangle = \hat{A}_j^{\dagger}\hat{A}_j^{\phantom\dagger}|\phi_{j-1}\rangle=0 \label{eq: SFM_6}
\end{equation}
since $\hat{A}_j^{\phantom\dagger}|\phi_{j-1}\rangle=0$. Equation~(\ref{eq: SFM_6}) then implies that $|\phi_{j-1}\rangle$ is an eigenstate of $\hat{\mathcal{H}}_j$ with eigenvalue $E_j$. The eigenvector of $\hat{\mathcal{H}}$ is next constructed via
\begin{equation}
    |\psi_j\rangle = \hat{A}_0^{\dagger}\hat{A}_1^{\dagger}...\hat{A}_{j-1}^{\dagger}|\phi_{j-1}\rangle.
    \label{eq: SFM_eigenstate}
\end{equation}
Using Eq.~(\ref{eq: SFM_6}) and repeated application of the Hermitian conjugation of Eq.~(\ref{eq: SFM_2}), we find
\begin{eqnarray}
\hat{\mathcal{H}}|\psi_j\rangle &=& \hat{\mathcal{H}}_0\hat{A}_0^{\dagger}\hat{A}_1^{\dagger}...\hat{A}_{j-1}^{\dagger}|\phi_{j-1}\rangle\nonumber \\
&=& \hat{A}_0^{\dagger}\hat{\mathcal{H}}_1\hat{A}_1^{\dagger}...\hat{A}_{j-1}^{\dagger}|\phi_{j-1}\rangle\nonumber \\
&=& \hat{A}_0^{\dagger}\hat{A}_1^{\dagger}...\hat{A}_{j-1}^{\dagger}\hat{\mathcal{H}}_j|\phi_{j-1}\rangle \nonumber\\
&=& \hat{A}_0^{\dagger}\hat{A}_1^{\dagger}...\hat{A}_{j-1}^{\dagger}E_j|\phi_{j-1}\rangle\nonumber \\
&=& E_j\hat{A}_0^{\dagger}\hat{A}_1^{\dagger}...\hat{A}_{j-1}^{\dagger}|\phi_{j-1}\rangle \nonumber\\
&=& E_j|\psi_j\rangle.
\end{eqnarray}
So $|\psi_j\rangle=\hat{A}_0^{\dagger}\hat{A}_1^{\dagger}...\hat{A}_{j-1}^{\dagger}|\phi_{j-1}\rangle$ is the eigenstate of $\hat{\mathcal{H}}$ with eigenvalue $E_j$!

While this discussion of the factorization method is abstract, we make it more concrete by showing how to determine the specific $\hat{A}_j^{\phantom\dagger}$, $\hat{A}_j^{\dagger}$, and $E_j$. The strategy is to pick
\begin{equation}
    \hat{A}_j^{\phantom\dagger} = \frac{\hat{p}}{\sqrt{2m}}+\frac{i\hbar}{\sqrt{2m}}k_jW_j(k_j'\hat{x})
\end{equation}
where $W_j$ is a real-valued function called the superpotential, and $k_j$ and $k_j'$ are real ``wavenumbers" with dimensions of inverse length. The form of this choice with constants $k_j$ and $k_j'$ may seem odd, but this is the form that is most useful for so-called shape-invariant potentials, because the auxiliary Hamiltonians can be constructed simply by choosing different values for these constants. We do not go into any further details on this, because the harmonic oscillator is the absolute simplest case, as we will see below.

Evaluating the product of the two operators next yields
\begin{equation}
    \hat{A}_j^{\dagger}\hat{A}_j^{\phantom\dagger} = \frac{\hat{p}^2}{2m} + i\frac{\hbar k_j}{2m}\left[ \hat{p}, W_j(k_j'\hat{x})\right] + \frac{\hbar^2 k_j^2}{2m}W_j^2(k_j'\hat{x}). \label{eq: SFM_7}
\end{equation}
Noting that $\hat{p}^2/2m$ denotes the kinetic-energy term, the rest of Eq.~(\ref{eq: SFM_7}) is the potential-energy contribution to the Hamiltonian $\hat{\mathcal{H}}_j$, minus a constant. From the definition of the $\hat{\mathcal{H}}_j$ given in Eq.~(\ref{eq: auxiliaryH}), we find that the missing constant term is just $E_j$. We then see that, for $j=0$, we have
\begin{equation}
    V(\hat{x}) = i\frac{\hbar k_0}{2m}\left[ \hat{p}, W_0(k_0'\hat{x})\right] + \frac{\hbar^2 k_0^2}{2m}W_0^2(k_0'\hat{x}) + E_0. \label{eq: SFM_8a}
\end{equation}
The entire challenge in this method lies in selecting the appropriate $W_j$. 

But for the simple harmonic oscillator, it is actually rather easy to find the superpotentials.
First note that Eq.~(\ref{eq: SFM_8a}) becomes
\begin{equation}
    \frac{1}{2}m\omega_0^2\hat{x}^2 = i\frac{\hbar k_0}{2m}\left[ \hat{p}, W_0(k_0'\hat{x})\right] + \frac{\hbar^2 k_0^2}{2m}W_0^2(k_0'\hat{x}) + E_0.
\end{equation}
For the simple harmonic oscillator, we choose $W_0(k_0'\hat{x}) = k_0'\hat{x}$,  use the canonical commutation relation to evaluate the commutator of the momentum with the superpotential, and set $k_0k_0'=-m\omega_0/\hbar$ (which also determines $E_0=\hbar\omega_0/2$). With these choices, we have
\begin{eqnarray}
\hat{A}_0^{\phantom\dagger} &=& \frac{1}{\sqrt{2m}}(\hat{p}-im\omega_0\hat{x}) \label{eq: SFM_8} \\ 
\hat{A}_0^{\dagger} &=& \frac{1}{\sqrt{2m}}(\hat{p}+im\omega_0\hat{x}), \label{eq: SFM_9}
\end{eqnarray}
which matches the notation for the ladder operator method of the simple harmonic oscillator given in many early quantum textbooks. However, the method and notation for the algebraic solution to the harmonic oscillator differs somewhat in today's texts. The abstract method was first introduced in the 1930 edition of Dirac's textbook on quantum mechanics\cite{Dirac30} (first edition) and further developed in his 1947 edition\cite{Dirac47} (third edition)  (more details of the history are provided below). The framework for the operator method has remained unchanged, but a different notation has since been universally adopted by quantum textbooks. The $i$ factors are moved from the coordinate to the momentum, and we work with dimensionless $\hat{a}$ and $\hat{a}^{\dagger}$ rather than the Schr\"odinger operators. The dimensionless (Dirac) ladder operators are then defined as
\begin{equation}
\hat a^\dagger = \sqrt{\frac{m\omega_0}{2\hbar}}\left (\hat x-i\frac{\hat p}{m\omega_0}\right ),~~
\hat a=\sqrt{\frac{m\omega_0}{2\hbar}}\left (\hat x+i\frac{\hat p}{m\omega_0}\right ).
\end{equation}
These operators differ by a factor of  $\pm i/\sqrt{\hbar\omega_0}$ from the corresponding Schr\"odinger operators given in Eqs.~(\ref{eq: SFM_8a}) and (\ref{eq: SFM_9}). We work now with the modern Dirac form of these operators due to their familiarity.

Our next task is to establish the eigenfunctions and eigenvalues of the simple harmonic oscillator following the Schr\"odinger approach. This methodology is different from Dirac's 1947 approach, which relies too heavily on the matrix mechanics approach, exploiting the raising and lowering operators to move up and down the spectrum. It is more closely aligned with the approach of Ikenberry,\cite{ikenberry} which employs instead the 1940 Schr\"odinger notion of positivity as the critical criterion for determining eigenstates after factorizing a Hamiltonian. Here is how it is done.

The (Dirac) raising and lowering operators satisfy
\begin{equation}
[\hat a,\hat a^\dagger]=\frac{m\omega_0}{2\hbar}\frac{i}{m\omega_0}2[\hat p,\hat x]=1, \label{eq: ham5}
\end{equation}
and 
\begin{equation}
\hat{\mathcal H}=\hbar\omega_0\left (\hat a^\dagger \hat a+\frac{1}{2}\right ).
\label{eq: ham2}
\end{equation}
Since $\hat a^\dagger\hat a$ is a positive semidefinite operator, it satisfies
\begin{equation}
\langle \psi|\hat a^\dagger \hat a|\psi\rangle=\lVert\hat a|\psi\rangle\rVert^2\ge 0
\end{equation}
for any state vector $|\psi\rangle$. Hence,
we learn that the ground state $|0\rangle$ of the simple harmonic oscillator requires
\begin{equation}
\hat a|0\rangle=0,
\end{equation}
and the ground-state energy is $E_0=\hbar\omega_0/2$.

We next find the relevant intertwining relationship: we operate $\hat{a}^{\dagger}$ on the right side of Eq.~(\ref{eq: ham2}) and discover that
\begin{eqnarray}
\hat{\mathcal{H}}\hat{a}^{\dagger} &=& \hbar\omega_0\left(\hat{a}^{\dagger}\hat{a}+\frac{1}{2}\right)\hat{a}^{\dagger}=\hbar\omega_0 \hat a^\dagger \left ( \hat a\hat a^\dagger +\frac{1}{2}\right ) \nonumber \\
&=& \hat{a}^{\dagger}\left[ \hat{\mathcal{H}}+\hbar\omega_0\right], \label{eq: ham3}
\end{eqnarray}
where the last line follows by applying the commutation relation of the Dirac operators. Hence, again recalling the Hermitian conjugate of Eq. (\ref{eq: SFM_2}), we have that all of the auxiliary Hamiltonians are just shifted by integer multiples of $\hbar\omega_0$, which also implies that $\hat{A}_j=\hat{A}_0 = -i\sqrt{\hbar\omega_0}\,\hat{a}$ for all $j$. Taking into account the different normalization of the Schr\"odinger and Dirac ladder operators, we then immediately find that
the eigenstates satisfy
\begin{equation}
    |n\rangle=\frac{\left (\hat a^\dagger\right )^n}{\sqrt{n!}}|0\rangle,
\end{equation}
with energies 
\begin{equation}
E_n=\hbar\omega_0\left (n+\frac{1}{2}\right ).
\end{equation}
This derivation uses Eqs.~(\ref{eq: SFM_5}) and (\ref{eq: SFM_eigenstate}). It also uses the normalization factor which relates the Schr\"odinger and Dirac ladder operators. Finally, we assume the ground state $|0\rangle$ is normalized from the beginning ($\langle0|0\rangle=1$).
This derivation differs from the standard approach, but we think it works better logically since it first determines the ground state from the factorization and a positivity argument and then constructs the excited states directly from the intertwining relation. Normalization then follows as the last step.

\section{Algebraic derivation of the wavefunctions of the simple harmonic oscillator}

We begin the algebraic derivation of the wavefunctions by simply noting that they are the overlap of the energy eigenfunctions $|n\rangle$ with the position $|x\rangle$ and momentum $|p\rangle$ eigenfunctions, or $\psi_n(x)=\langle x|n\rangle$ and $\phi_n(p)=\langle p|n\rangle$. So we must start by constructing those position and momentum eigenfunctions. Our strategy is to employ operator methods without resorting to specific representations of the operators, so we do not need to  introduce the  coordinate-space representation of the momentum operator in terms of a derivative with respect to the position. Instead, we follow the representation-independent operator-based approach initiated by Pauli\cite{pauli_1925} and independently by Dirac\cite{dirac_1926} in 1926.

 Our goal is to use these operators to construct position eigenstates. We will assume that an eigenstate for  position at the origin exists and is denoted $|x{=}0\rangle$. It satisfies $\hat x|x{=}0\rangle=0$, and use that state to construct all other position eigenstates. Note that we do not need to worry about the normalization of the state for anything that we do here, so we do not discuss this issue further (as its treatment is well covered in all quantum texts).

We will employ the Hadamard lemma, which is given by
\begin{equation}
e^{\hat A}\hat Be^{-\hat A}=\hat B+\sum_{m=1}^\infty\frac{1}{m!}[\hat A,[\hat A,\cdots ,[\hat A,\hat B]\cdots]_m
\label{eq: hadamard}
\end{equation}
where the $m$ subscript on the commutators denotes that there are $m$ nested commutators; this lemma is also called the Baker-Hausdorff lemma and the braiding relation. But as far as we can tell, it was first discovered by Campbell in 1897 [see Eq.~(19) of the historical discussion of the Baker-Campbell-Hausdorff relation\cite{bch_history}] and hence should be called the Campbell lemma. Despite significant research, we were unable to determine where the Hadamard lemma name comes from.

Before we jump into the derivation of position and momentum operators, we note that the Hadamard lemma can be employed to establish some additional identities. Any function $f(\hat B)$ of an operator $\hat B$ that can be written as a power series in $\hat B$ satisfies
\begin{eqnarray}
 e^{\hat A}f(\hat B)e^{-\hat A}&=&e^{\hat A}\sum_{m=0}^\infty f_m\hat B^me^{-\hat A}=\sum_{m=0}^\infty f_m\left (e^{\hat A}\hat Be^{-\hat A}\right )^m\nonumber\\
&=&f(e^{\hat A}\hat Be^{-\hat A})\nonumber\\
&=&f\left (\sum_{m=0}^\infty\frac{1}{m!}[\hat A,[\hat A,\cdots ,[\hat A,\hat B]\cdots]_m\right ).
\label{eq: braiding_general}
\end{eqnarray}
This is an exact relation. Choosing $f(\hat B)=\exp(\hat B)$ then yields an important identity after some simple re-arranging of terms:
\begin{equation}
e^{\hat A}e^{\hat B}=e^{\sum_{m=0}^\infty\frac{1}{m!}[\hat A,[\hat A,\cdots ,[\hat A,\hat B]\cdots]_m}e^{\hat A}.
\end{equation}
This relation is often called the braiding relation.
When $[\hat A,\hat B]$ commutes with $\hat A$ and $\hat B$, we then have the exponential re-ordering identity
\begin{equation}
e^{\hat A}e^{\hat B}=e^{\hat B}e^{\hat A}e^{[\hat A,\hat B]},
\label{eq: braiding_simple}
\end{equation}
which includes a correction term when the exponential operators are re-ordered.

To start working with the translation operator we use the Hadamard lemma in Eq.~(\ref{eq: hadamard}), which allows us to evaluate the similarity transformation of the operator $\hat x$ as follows (with $x_0$ a real number):
\begin{eqnarray}
e^{\frac{i}{\hbar}x_0\hat p}\hat xe^{-\frac{i}{\hbar}x_0\hat p}&=&\hat x+\frac{i}{\hbar}x_0[\hat p,\hat x]-\frac{x_0^2}{2\hbar^2}[\hat p,[\hat p,\hat x]]+\cdots\nonumber\\
&=&\hat x+x_0.
\label{eq: trans1}
\end{eqnarray}
The final equality occurs because $[\hat p,\hat x]=-i\hbar$ is a number, not an operator, and subsequently it commutes with all additional multiple commutators of $\hat p$. This truncates the Hadamard lemma expression after the first commutator. Next, we multiply both sides of Eq.~(\ref{eq: trans1}) by $\exp(-ix_0\hat p/\hbar)$ from the left to yield
\begin{equation}
\hat xe^{-\frac{i}{\hbar}x_0\hat p}=e^{-\frac{i}{\hbar}x_0\hat p}(\hat x+x_0).
\end{equation}
With this identity, we establish the eigenfunction $|x_0\rangle$, which satisfies $\hat x|x_0\rangle=x_0|x_0\rangle$ (here, $x_0$ is a number and a label for the Dirac ket):
\begin{equation}
|x_0\rangle=e^{-\frac{i}{\hbar}x_0\hat p}|x{=}0\rangle.
\label{eq: trans2}
\end{equation}
Operating $\hat x$ onto the state $|x_0\rangle$ yields
\begin{equation}
\hat x|x_0\rangle=\hat xe^{-\frac{i}{\hbar}x_0\hat p}|x{=}0\rangle=
e^{-\frac{i}{\hbar}x_0\hat p}(\hat x+x_0)|x{=}0\rangle=x_0|x_0\rangle.
\label{eq: trans3}
\end{equation}
The last equality follows from $\hat x|x{=}0\rangle=0$, the fact that numbers always commute with operators and the definition of $|x_0\rangle$. Hence, Eqs.~(\ref{eq: trans2}) and (\ref{eq: trans3}) establish that $|x_0\rangle$ is an eigenstate of $\hat x$ with eigenvalue $x_0$.

Similarly, one can also derive that the momentum eigenstates satisfy
\begin{equation}
|p_0\rangle=e^{\frac{i}{\hbar}p_0\hat x}|p{=}0\rangle.
\end{equation}
Note the different sign in the exponent for the position and momentum eigenfunctions.
This result will be used in subsequent calculations.

We are almost ready to compute the coordinate-space wavefunction using purely algebraic methods. The derivation requires one more identity: the Baker-Campbell-Hausdorff (BCH) identity.\cite{baker,campbell,hausdorff}
The BCH identity is ``halfway'' between the exponential re-ordering identity, which rewrites the exponential of the sum of the operators in terms of the two exponential operators and a correction factor---here, the BCH formula takes a product of exponential of operators and rewrites it as the exponential of a new operator. Unlike the Hadamard lemma and its application to exponential re-ordering, the BCH identity does not have any simple explicit formula for its result in the general case (although one can write the result in closed form\cite{dynkin,bch_closed}). Fortunately for us, we need it only for the case where $[\hat A,\hat B]$ commutes with $\hat A$ and $\hat B$---in this case, the BCH result greatly simplifies and is given by
\begin{eqnarray}
    e^{\hat{A}}e^{\hat{B}}&=&e^{\hat{A}+\hat{B}+\frac{1}{2}[\hat{A},\hat{B}]} \label{eq: BCH_fin1}
\\
    e^{\hat{B}}e^{\hat{A}}&=&e^{\hat{A}+\hat{B}-\frac{1}{2}[\hat{A},\hat{B}]} \label{eq: BCH_fin2}
\end{eqnarray}
The BCH identity is a well-known and well-established result, so we do not provide its derivation here.

We now have all the technical tools needed to determine the coordinate-space wavefunction $\psi_n(x)=\langle x|n\rangle$. Using the position eigenstates and the energy eigenstates, we immediately find that
\begin{equation}
\psi_n(x)=\langle x|n\rangle=\frac{1}{\sqrt{n!}}\langle x{=}0|e^{\frac{i}{\hbar}x\hat p}\left (\hat a^\dagger\right )^n|n{=}0\rangle.
\end{equation}
The operators $\hat p$ and $\hat a^\dagger$ can be easily identified by their hats. Note that one can think of this representation in the following way: at the origin, the wavefunction is $\psi_n(0)=\langle x{=}0|n\rangle$ (which is a number that will ultimately be fixed by normalization) and the translation operator then shifts the wavefunction from the origin to the position $x$ and tells us how the wavefunction value changes in the process. This allows us to compute the wavefunction everywhere by shifting the value of the coordinate. The algebraic computation then simply evaluates the operator expression.

The strategy to determine the wavefunction algebraically now takes a few additional steps. First, we replace the momentum operator in the exponent of the translation operator by its expression in terms of the ladder operators
\begin{equation}
\hat p=-i\sqrt{\frac{m\hbar\omega_0}{2}}\left (\hat a-\hat a^\dagger\right ).
\end{equation}
The wavefunction becomes
\begin{equation}
\psi_n(x)=\frac{1}{\sqrt{n!}}\langle x{=}0|e^{\sqrt{\frac{m\omega_0}{2\hbar}}x(\hat a-\hat a^\dagger)}\left (\hat a^\dagger\right )^n|n{=}0\rangle.
\end{equation}
Then we use the BCH relation in Eq.~(\ref{eq: BCH_fin1}) with $\hat A\propto \hat a^\dagger$ and $\hat B\propto \hat a$ to factorize the translation operator into a factor involving the raising operator on the left and the lowering operator on the right. This is given by
\begin{equation}
\psi_n(x)=\frac{1}{\sqrt{n!}}e^{-\frac{m\omega_0}{4\hbar}x^2}\langle x{=}0|e^{-\sqrt{\frac{m\omega_0}{2\hbar}}x\hat a^\dagger}e^{\sqrt{\frac{m\omega_0}{2\hbar}}x\hat a}\left (\hat a^\dagger\right )^n|n{=}0\rangle.
\end{equation}
Third, we take the relation in Eq.~(\ref{eq: braiding_general}) and multiply by $\exp(\hat A)$ on the right to create the general functional braiding relation and apply it to the matrix element for the wavefunction with $f(\hat B)=(\hat a^\dagger)^n$. This yields
\begin{equation}
\psi_n(x)=\frac{1}{\sqrt{n!}}e^{-\frac{m\omega_0}{4\hbar}x^2} x{=}0|e^{-\sqrt{\frac{m\omega_0}{2\hbar}}x\hat a^\dagger}\left (\hat a^\dagger+\sqrt{\frac{m\omega_0}{2\hbar}}x\right )^ne^{\sqrt{\frac{m\omega_0}{2\hbar}}x\hat a}|n{=}0\rangle.
\end{equation}
The rightmost exponential factor gives 1 when it operates on the state because $\hat a|n{=}0\rangle=0$. Thus, we have
\begin{equation}
\psi_n(x)=\frac{1}{\sqrt{n!}}e^{-\frac{m\omega_0}{4\hbar}x^2}\langle x{=}0|e^{-\sqrt{\frac{m\omega_0}{2\hbar}}x\hat a^\dagger}\left (\hat a^\dagger+\sqrt{\frac{m\omega_0}{2\hbar}}x\right )^n|n{=}0\rangle.
\end{equation}
Next, we introduce a new exponential factor with the opposite sign of the exponent multiplying the ground-state wavefunction, because it equals 1 when operating against the state:
\begin{equation}
\psi_n(x)=\frac{1}{\sqrt{n!}}e^{-\frac{m\omega_0}{4\hbar}x^2}\langle x{=}0|e^{-\sqrt{\frac{m\omega_0}{2\hbar}}x\hat a^\dagger}\left (\hat a^\dagger+\sqrt{\frac{m\omega_0}{2\hbar}}x\right )^ne^{-\sqrt{\frac{m\omega_0}{2\hbar}}x\hat a}|n{=}0\rangle.
\end{equation}
The general functional braiding relation is used again to bring the rightmost exponential factor to the left through the $\hat a^\dagger$ term raised to the $n$th power
\begin{equation}
\psi_n(x)=\frac{1}{\sqrt{n!}}e^{-\frac{m\omega_0}{4\hbar}x^2}\langle x{=}0|e^{-\sqrt{\frac{m\omega_0}{2\hbar}}x\hat a^\dagger}e^{-\sqrt{\frac{m\omega_0}{2\hbar}}x\hat a}\left (\hat a^\dagger+\sqrt{\frac{2m\omega_0}{\hbar}}x\right )^n|n{=}0\rangle.
\end{equation}
Now, we use the BCH relation again to combine the two exponentials into one which increases the Gaussian exponent by a factor of two
\begin{equation}
\psi_n(x)=\frac{1}{\sqrt{n!}}e^{-\frac{m\omega_0}{2\hbar}x^2}\langle x{=}0|e^{-\sqrt{\frac{m\omega_0}{2\hbar}}x(\hat a^\dagger+\hat a)}\left (\hat a^\dagger+\sqrt{\frac{2m\omega_0}{\hbar}}x\right )^n|n{=}0\rangle.
\end{equation}
Finally, we use the fact that the sum of the raising and lowering operator is proportional to the position operator
\begin{equation}
\hat x=\sqrt{\frac{\hbar}{2m\omega_0}}\left (\hat a+\hat a^\dagger\right ).
\end{equation}
We replace the sum of the raising and lowering operator in the exponent and let it act on the state to the left, where it gives 1, because the position operator annihilates the state $\langle x{=}0|$. The wavefunction has now become
\begin{equation}
\psi_n(x)=\frac{1}{\sqrt{n!}}e^{-\frac{m\omega_0}{2\hbar}x^2}\langle x{=}0|\left (\hat a^\dagger+\sqrt{\frac{2m\omega_0}{\hbar}}x\right )^n|n{=}0\rangle.
\end{equation}

We are almost done. We have achieved a reduction of the problem into a Gaussian function multiplied by a matrix element which is an $n$th degree polynomial in $x$. All that is left is evaluating the polynomial. To do this, we first introduce a definition of the polynomial, which we will then show is a so-called Hermite polynomial $H_n$. We write the wavefunction as 
\begin{equation}
\psi_n(x)=\frac{1}{\sqrt{n!2^n}}H_n\left (\sqrt{\frac{m\omega_0}{\hbar}}x\right )e^{-\frac{m\omega_0}{2\hbar}x^2}\langle x{=}0|n{=}0\rangle,
\end{equation}
which defines the Hermite polynomial via
\begin{equation}
H_n\left (\sqrt{\frac{m\omega_0}{\hbar}}x\right )=\frac{\sqrt{2^n}}{\langle x{=}0|n{=}0\rangle}\langle x{=}0|\left (\hat a^\dagger+\sqrt{\frac{2m\omega_0}{\hbar}}x\right )^n|n{=}0\rangle.
\label{eq: hermite}
\end{equation}
Note that the number $\langle x{=}0|n{=}0\rangle$ is the normalization constant for the ground-state wavefunction; we will discuss how to determine it below. This definition allows us to immediately determine the first two polynomials $H_0$ and $H_1$. Choosing $n=0$ in Eq.~(\ref{eq: hermite}) immediately yields $H_0=1$. Choosing $n=1$, produces
\begin{equation}
H_1\left (\sqrt{\frac{m\omega_0}{\hbar}}x\right )=2\sqrt{\frac{m\omega_0}{\hbar}}x+\frac{\sqrt{2}}{\langle x{=}0|n{=}0\rangle}\langle x{=}0|\hat a^\dagger|n{=}0\rangle.
\end{equation}
The second term vanishes for the following reason: we first note that $\hat a^\dagger|n{=}0\rangle=(\hat a^\dagger+\hat a)|n{=}0\rangle$, because the lowering operator annihilates the ground state. Hence $\hat a^\dagger|n{=}0\rangle\propto \hat x|n{=}0\rangle$. But $\langle x{=}0|\hat x=0$, so this state vanishes when it acts against the position eigenstate. 

For the remainder of the Hermite polynomials, we work out a two-term recurrence relation. We focus on the nontrivial matrix element, and factorize the terms as follows:
\begin{equation}
\langle x{=}0|\left (\hat a^\dagger+\sqrt{\frac{2m\omega_0}{\hbar}}x\right )
\left (\hat a^\dagger+\sqrt{\frac{2m\omega_0}{\hbar}}x\right )^{n-1}|n{=}0\rangle.
\end{equation}
The constant term in the first factor can be removed from the matrix element and it multiplies the matrix element with $n-1$ operator factors (which is proportional to $H_{n-1}$). For the remaining term proportional to $\hat a^\dagger$, we replace the operator by $\hat a^\dagger\to \hat a^\dagger+\hat a-\hat a$. The term proportional to $\hat a^\dagger +\hat a$ is proportional to $\hat x$, and so it annihilates when it operates on the left against the $\langle x{=}0|$ state. The remaining $\hat a$ operator can be replaced by the commutator of the $n-1$ power of the $\hat a^\dagger$ term, because $\hat a|n{=}0\rangle=0$. Generalizing the standard result $[\hat a,(\hat a^\dagger)^n]=n(\hat a^\dagger)^{n-1}$, the remaining commutator is straightforward to evaluate via
\begin{equation}
\left [ \hat a,\left (\hat a^\dagger+\sqrt{\frac{2m\omega_0}{\hbar}}x\right )^{n-1}\right ]=
(n-1)\left (\hat a^\dagger+\sqrt{\frac{2m\omega_0}{\hbar}}x\right )^{n-2}.
\end{equation}
We can assemble all of these results to find the recurrence relation for the Hermite polynomials, which becomes
\begin{equation}
H_n\left (\sqrt{\frac{m\omega_0}{\hbar}}x\right )=2\sqrt{\frac{m\omega_0}{\hbar}}xH_{n-1}\left (\sqrt{\frac{m\omega_0}{\hbar}}x\right )-2(n-1)H_{n-2}\left (\sqrt{\frac{m\omega_0}{\hbar}}x\right ).
\end{equation}
This recurrence relation, which is of the form $H_n(z)=2zH_{n-1}(z)-2(n-1)H_{n-2}(z)$, is the standard Hermite polynomial recurrence relation when $H_0(z)=1$ and $H_1(z)=2z$, as we have here.

We have now established that the simple-harmonic-oscillator wavefunction satisfies
\begin{equation}
\psi_n(x)=\frac{1}{\sqrt{n!2^n}}H_n\left (\sqrt{\frac{m\omega_0}{\hbar}}x\right )e^{-\frac{m\omega_0}{2\hbar}x^2}\langle x{=}0|n{=}0\rangle.
\end{equation}
The last task in front of us is to find the normalization factor. This is computed for the ground state via
\begin{equation}
|\langle x{=}0|n{=}0\rangle|^2\int_{-\infty}^\infty dx e^{-\frac{m\omega_0}{\hbar}x^2}=1
\end{equation}
or
\begin{equation}
\langle x{=}0|n{=}0\rangle=\left ( \frac{m\omega_0}{\pi\hbar}\right )^{\frac{1}{4}}.
\end{equation}
We have finally produced the wavefunction for the simple harmonic oscillator using algebraic methods. Note that calculus is only needed for the last normalization step.

We end this section with a brief sketch of how one uses similar methods to determine the momentum-space wavefunctions. To start, the momentum ``boost'' operator is given by $\exp(ip\hat x/\hbar)$, and the momentum eigenstates satisfy
\begin{equation}
|p\rangle=e^{\frac{i}{\hbar}p\hat x}|p{=}0\rangle.
\end{equation}
The wavefunction is given by $\phi_n(p)=(i)^n\langle p|n\rangle$; we added an additional global phase to ensure we reproduce the standard results---you will see why this is important below. The wavefunction can be expressed in terms of the operators as
\begin{equation}
\phi_n(p)=\frac{(i)^n}{\sqrt{n!}}\langle p{=}0|e^{-\frac{i}{\hbar}p\hat x}\left (\hat a^\dagger\right )^n|n{=}0\rangle.
\end{equation}
The remainder of the calculations proceeds as before for the coordinate-space wavefunction. We start by replacing the $\hat x$ operator by the sum of raising and lowering operators; in this case, the coefficients of the raising and lowering operators are now purely imaginary. We use BCH to factorize the exponential into a raising operator on the left and lowering operator on the right. Then we use the braiding identity to move the exponential through the $(\hat a^\dagger)^n$ terms and let it operate on the ground state, where it produces 1. The shift term added to the raising operator is now purely imaginary. Next, we introduce a factor of 1 at the ground state, which is the same exponential operator of the lowering operator but with the sign of the exponent changed. Then we use the braiding identity to bring it back to the left, BCH to place the operators in one exponential, and evaluate the momentum operator on the momentum eigenstate. At this stage, the wavefunction has become
\begin{equation}
\phi_n(p)=\frac{(i)^n}{\sqrt{n!}}e^{-\frac{p^2}{2\hbar\omega_0m}}\langle p{=}0|\left (\hat a^\dagger-i\frac{\sqrt{2}p}{\sqrt{\hbar\omega_0m}}\right )^n|n{=}0\rangle.
\end{equation}
Note the additional factors of $i$ and the replacement of $\sqrt{m\omega_0/\hbar}x$ by $p/\sqrt{\hbar\omega_0 m}$. The Hermite polynomial now needs to be defined via
\begin{equation}
H_n\left (\frac{p}{\sqrt{\hbar\omega_0m}}\right )=\frac{\sqrt{2^n}i^n}{\langle p{=}0|n{=}0\rangle}\langle p{=}0|\left (\hat a^\dagger-i\frac{\sqrt{2}p}{\sqrt{\hbar\omega_0m}}\right )^n|n{=}0\rangle.
\end{equation}
Starting with $H_0=1$ and $H_1=2p/\sqrt{\hbar\omega_0m}$, we find the same Hermite polynomials as we found before, but now with $z=p/\sqrt{\hbar\omega_0m}$. The rest of the calculation is similar to the coordinate space calculation.
The normalization factor is found by a simple integral. One can see that this procedure will lead to the momentum-space wavefunction, which finally satisfies
\begin{equation}
\phi_n(p)=\frac{1}{(\pi\hbar\omega_0m)^{\frac{1}{4}}}\frac{1}{\sqrt{n!2^n}}H_n\left (\frac{p}{\sqrt{\hbar\omega_0 m}}\right )e^{-\frac{p^2}{2\hbar\omega_0m}}.
\end{equation}
Aside from some different constants, the coordinate-space and momentum-space  wavefunctions have identical functional forms. This is expected from the outset, because the Hamiltonian is quadratic in both momentum {\it and} position. Hence, the wavefunctions must be isomorphic.

This ends our algebraic derivation of the wavefunctions of the simple harmonic oscillator. We hope that you will try employing it the next time you teach a quantum mechanics class.

\section{History of the Simple Harmonic Oscillator in Quantum Mechanics}

We now look into the historical development of the operator method for the simple harmonic oscillator. Although much work has been done on the history of quantum mechanics, it seems no one has attempted an in-depth exploration of the harmonic oscillator. There is no mention in standard quantum historical texts, including Jammer,\cite{Jammer} Taketani and Nagasaki's\cite{TaketaniNagasaki} three-volume work, and even  Mehra and Rechenberg's\cite{MehraRechenberg} six-volume set on the history of quantum mechanics. In his discussion of transformation theory, Purrington\cite{Purrington} does mention the introduction of ladder operators for the harmonic oscillator in Born and Jordan's textbook.\cite{BornJordan30}~ However, our interpretation of Born and Jordan's book differs from that of Purrington, as we read the Born and Jordan text as working with Heisenberg matrices of the raising and lowering operators. Thus, we don't consider their approach an abstract operator formalism. While the aforementioned texts expound on the evolution of a variety of areas in quantum mechanics, none of them trace the progression of the solutions of the harmonic oscillator. One explanation for this might be a simple lack of interest in the harmonic oscillator during the early development of quantum theory. Most of the original publications that developed quantum mechanics in the period from 1925-30 were primarily interested in determining the atomic spectra of elements other than Hydrogen and in quantizing light. In addition, the simple harmonic oscillator spectrum was determined in the first matrix mechanics papers by Heisenberg\cite{Heisenberg25} and Born and Jordan.\cite{BornJordan25}~ Schr\"odinger solved it in his second paper,\cite{Schrodinger26} providing both the spectrum and the wavefunctions (via a differential equations approach). So, the harmonic oscillator seems to have slipped through the cracks, and its historical study remains underdeveloped. Starting from the 1920s, we seek here to provide an understanding of the development of the quantum-mechanical solutions of the simple harmonic oscillator. Note that from time to time we will use the original notation employed in the original articles. We try to make it clear when this is being done below.

Heisenberg was the first to find the energies of the harmonic oscillator in his 1925 paper\cite{Heisenberg25} that invented modern quantum mechanics. His seminal paper relied on classical equations of motion and replaced them with their matrix-valued quantum counterparts (a strategy similar to the old quantum mechanics method of Bohr-Sommerfeld quantization). Using this matrix-valued equation of motion and the canonical commutation relation, Heisenberg is able to find the quantized energy levels. The first problem treated was that of an anharmonic oscillator with a third-order perturbation term. Heisenberg truncates his result to determine the energies for the unperturbed harmonic oscillator:
\begin{equation}
W = \hbar\omega_0\left( n + \frac{1}{2} \right). \label{eq: Heisenberg}
\end{equation}
While Heisenberg's article provides essentially no details for how the calculation was done,\cite{Aitchison} he does compute the correct result. Born and Jordan published a paper\cite{BornJordan25} shortly after Heisenberg's in which they provide the details of the matrix-mechanics solution for the simple harmonic oscillator. The matrix mechanics methodology does contain many elements of the operator method which Dirac later developed in the first three editions of his textbook.\cite{Dirac30, Dirac35, Dirac47}~ Matrix mechanics works by essentially determining properties of the position space matrix, defined in modern terms via
\begin{equation}
    q_{mn}(t)=\langle m|e^{\frac{i}{\hbar}{\mathcal H}t}\hat q e^{-\frac{i}{\hbar}{\mathcal H}t}|n\rangle.
\end{equation}
One can see that the time-dependence of the matrix goes like $\exp[-i(E_n-E_m)t/\hbar]$. Substituting into the classical equation of motion for the simple harmonic oscillator yields the constraint that $E_n-E_m=\pm \hbar\omega_0$. Hence, the $\hat q$ matrix is tridiagonal, and the consecutive energy levels are separated in steps of $\hbar\omega_0$. Next, positivity of the Hamiltonian is used to show that there must exist some minimum energy level equal to $\frac{1}{2}\hbar\omega_0$. From this ladder of energies, they deduce that the $n$th diagonal value of the Hamiltonian is given by Heisenberg's result in Eq.~(\ref{eq: Heisenberg}). The connection between Born and Jordan's paper and the ladder operator method is further exhibited in Birtwistle's textbook\cite{Birtwhistle} which presents diagrams in a ladder formation connecting the different energy levels.

These matrix-mechanics papers failed to treat the eigenstates of the harmonic oscillator since matrix mechanics has no concept of an eigenfunction. It was not until Schr\"odinger introduced the wavefunction in 1926 that quantum papers began to explicitly refer to the eigenstates of the harmonic oscillator. In his paper,\cite{Schrodinger26} Schr\"odinger not only introduces the wavefunction but also develops the differential equation method for treating the harmonic oscillator. Using his time-independent wave equation for a harmonic potential
\begin{equation}
    \frac{d^2\psi_n(q)}{dq^2}+\frac{2m}{\hbar^2}\left (E_n-\frac{1}{2}m\omega_0^2q^2\right )\psi_n(q)=0,
\end{equation}
Schr\"odinger finds the energies of the harmonic oscillator as well as its eigenstates, which he expresses (unnormalized) in the coordinate-space representation as
\begin{equation}
\psi_n(q) = e^{-\frac{m\omega_0}{2\hbar^2}q^2}H_n\left(q\sqrt{\frac{\omega_0}{\hbar}}\right),
\end{equation}
where $H_n$ again denotes the Hermite polynomials. Schr\"odinger thus introduced the differential equation method now universally employed in all quantum textbooks, and his articulation of the eigenstate enabled the development of the operator method in early editions of Dirac's textbook.\cite{Dirac30, Dirac47}~ Dirac, like his contemporaries, discusses matrix mechanics in his 1930 textbook. Indeed, the relationship between matrix mechanics and operator methods is quite close.

Before jumping into the development of the ladder operator method for the harmonic oscillator, we must mention the appearance of bosonic creation and annihilation operators in other areas of quantum theory. As noted earlier, a principal concern of many early quantum papers was the quantization of light. Consequently, Dirac,\cite{Dirac27} Jordan,\cite{Jordan27} and Fock\cite{Fock32} all published papers in the late 1920s and early 1930s which include bosonic creation and annihilation operators. While at the time it appears that they were unaware of the relation between these operators and the harmonic oscillator, their publications coincide with the origins of the ladder operator method presented here. Since it was present in other areas of quantum theory at the time, we can see then that the notion of ladder operators was not unique to the early treatment of the harmonic oscillator.

We also mention one other item which was of great interest to the quantum pioneers---the theory of canonical transformations and the formulation of quantum mechanics in terms of action-angle variables. Here, Dirac led the way in his first quantum paper\cite{dirac25} on canonical quantization, where he nearly constructs the raising and lowering operators toward the end of the paper. He does note that the approach works for the simple harmonic oscillator but provides no details. Fritz London produced similar work in a 1926 paper,\cite{London27} although the raising and lowering operators do not explicitly appear in his work either.

The first work to formally define two operators which factorize the Hamiltonian of the harmonic oscillator is Born and Jordan's 1930 textbook,\cite{BornJordan30} which was completed a few months \textit{before} Dirac's first edition.\cite{Dirac30}~ They write the Hamiltonian as
\begin{equation}
    {\mathcal{H}} = \frac{1}{2\mu}{p}^2+\frac{a}{2}{q}^2
\end{equation}
where $\mu$ represents mass and $a$ what they call the quasi-elastic constant. Born and Jordan introduce two matrices
\begin{equation}
b = C(p-2\pi i \nu_0 \mu q) \quad{\rm and}\quad
b^{\dagger} = C(p+2\pi i \nu_0 \mu q) ,
\end{equation}
where $C = \frac{1}{\sqrt{2h\nu_0\mu}}$. They note that ${b}{b}^{\dagger}-{b}^{\dagger}{b}=1$ and re-write the Hamiltonian as
\begin{equation}
    {\mathcal{H}} = h\nu_0{b}{b}^{\dagger}-\frac{h\nu_0}{2} =h\nu_0{b}^{\dagger}{b}+\frac{h\nu_0}{2}.
\end{equation}
Born and Jordan's definition of $b$ and $b^{\dagger}$ and subsequent rewriting of the Hamiltonian appears nearly identical to the modern operator method (which instead uses $\hat{a}$ and $\hat{a}^{\dagger}$). Although they refer to them as ``Stufenmatrizen," Born and Jordan don't seem to use $b$ and $b^{\dagger}$ as ladder operators which act directly on eigenstates. We then do not consider this approach to be the initial formulation of the abstract operator method. Born and Jordan apparently wrote their 1930 textbook as a last-ditch-effort to save matrix mechanics from oblivion. This did not happen, and unfortunately the textbook has been nearly forgotten (in part because it was never translated into English).

The operator method for the simple harmonic oscillator then takes its first form in the 1930 edition\cite{Dirac30} of Dirac's textbook, although his discussion is quite similar to Born and Jordan's and inherits much of the matrix-mechanics argument. Dirac works with a dimensionless abstract Hamiltonian first. To find the eigenvalues of
\begin{equation}
\hat{\mathcal H}=\hat{p}^2+\hat{q}^2,
\end{equation}
Dirac defines an operator
\begin{equation}
\hat{A} = (\hat{p}+i\hat{q})(\hat{p}-i\hat{q}), \label{eq: dirac1}
\end{equation}
which a simple calculation shows to be essentially the Hamiltonian for the harmonic oscillator. He defines the eigenstates of $\hat{A}$ to satisfy the standard eigenvalue equation
\begin{equation}
\hat{A}| A' \rangle = A'|A'\rangle
\end{equation}
and then proceeds through a matrix-mechanics argument to show that
$\langle A' | (\hat{p}+i\hat{q}) | A'' \rangle$
equals zero unless $A'' = A'-2$. Using this and the non-negativity of $\hat{p}^2+\hat{q}^2$, Dirac finds that the eigenvalues of $\hat{A}$ are all the even non-negative integers: $0,2,4,6...$ and so on. From his earlier assertion that
\begin{equation}
\langle A' | (\hat{p}+i\hat{q}) | A'' \rangle = \delta_{A'',A'-2}
\end{equation}
we can then see how $(\hat{p}+i\hat{q})$ acts as a ladder operator on $|A''\rangle$ to raise it to the next highest eigenstate of $\hat{A}$. Dirac's expression given in Eq.~(\ref{eq: dirac1}) then shows that $\hat{A}$ is analogous to the ladder operator formulation of the Hamiltonian.
What Dirac's initial treatment lacked was a formulation of the eigenstate in terms of operators acting on the ground state (which we conjecture is because he adopted a matrix-mechanics methodology to find the spectrum and matrix-mechanics does not construct eigenstates). Dirac gives further allusion to the ladder operators by introducing their matrix representation
\arraycolsep=5pt\def\arraystretch{0.7}
\begin{equation}
\left(
\begin{array}{ccccccc}
0 & 0 & 0 & 0 & 0 & \cdot & \cdot \\
1 & 0 & 0 & 0 & 0 & \cdot & \cdot \\
0 & 1 & 0 & 0 & 0 & \cdot & \cdot \\
0 & 0 & 1 & 0 & 0 & \cdot & \cdot \\
0 & 0 & 0 & 1 & 0 & \cdot & \cdot \\
\cdot & \cdot & \cdot & \cdot & \cdot & \cdot & \cdot \\
\end{array}
\right)\quad{\rm and}\quad
\left(
\begin{array}{ccccccc}
0 & 1 & 0 & 0 & 0 & \cdot & \cdot \\
0 & 0 & 1 & 0 & 0 & \cdot & \cdot \\
0 & 0 & 0 & 1 & 0 & \cdot & \cdot \\
0 & 0 & 0 & 0 & 1 & \cdot & \cdot \\
0 & 0 & 0 & 0 & 0 & \cdot & \cdot \\
\cdot & \cdot & \cdot & \cdot & \cdot & \cdot & \cdot \\
\end{array}
\right), \label{eq: dirac2}
\end{equation}
which he denotes via the unconventional notation $e^{i\omega}$ and $e^{-i\omega}$, respectively. He notes that we can write the momentum and position operators as
\begin{eqnarray}
\hat{p}&=& \sqrt{\frac{m\omega}{2}}(\hat{J}^{\frac{1}{2}}e^{i\omega}+e^{-i\omega}\hat{J}^{\frac{1}{2}}) \label{eq: dirac6} \\
\hat{q} &=& \sqrt{\frac{1}{2m\omega}}(-i\hat{J}^{\frac{1}{2}}e^{i\omega}+ie^{-i\omega}\hat{J}^{\frac{1}{2}}), \label{eq: dirac7}
\end{eqnarray}
where $\hat{J}$ is denoted the ``action variable" and given by
\begin{equation}
\hat{J} = \frac{\hat{\mathcal{H}}}{\omega}-\frac{1}{2}\hbar\mathbb{I}. \label{eq: dirac3}
\end{equation}
With Eqs.~(\ref{eq: dirac2}) and (\ref{eq: dirac3}), we can calculate that
\begin{equation}
\hat{J}^{\frac{1}{2}}e^{i\omega} = \sqrt{\hbar}\hat{a}^{\dagger} \quad{\rm and}\quad
e^{-i\omega}\hat{J}^{\frac{1}{2}} = \sqrt{\hbar}\hat{a}, \label{eq: dirac4}
\end{equation}
where $\hat{a}$ and $\hat{a}^{\dagger}$ are the ladder operators commonly used to treat the harmonic oscillator today (but {\it not} introduced by Dirac in 1930). Furthermore, with Eqs.~(\ref{eq: dirac4}) above, we can also see that the form of Eqs.~(\ref{eq: dirac6}) and (\ref{eq: dirac7}) is almost identical to the way the momentum and position operators are defined today in terms of the ladder operators. Finally, Dirac's 1930 textbook seems to be the first to give the wavefunctions of the harmonic oscillator as the overlap of the energy eigenstates with position space, which he writes as an inner product $( q | n )$,
where $|n)$ denotes the $n$th eigenstate (this was written {\it before} Dirac notation was introduced). Dirac uses differential equations to find the wavefunctions, which he expresses with finite power series in $q$ (using the standard Frobenius series solution method). While the operator method in his 1930 textbook contains remarkable similarities to that in modern textbooks, there remain a few differences to point out. Dirac does not formally define the ladder operators here but instead uses expressions of the form $(\hat{p} \pm i\hat{q})$ as ladder operators---indeed, his approach presages the Schr\"odinger factorization method since it is focused on factorizing the Hamiltonian. We also note that Dirac includes the factor of $i$ on the position operator in Eqs. (\ref{eq: dirac1}) and (\ref{eq: dirac7}) above, which differs from the standard notation today, but again agrees with the Schr\"odinger factorization method. However, it is also fair to say that Dirac's approach is quite similar to the matrix mechanics methodology of Born and Jordan. Dirac uses the Heisenberg matrices to determine the eigenvalues in a standard matrix mechanics approach. His main difference is that he is the first to work with the operators by themselves instead of solely with the matrices (which is how we interpret the Born and Jordan methodology).

Other textbooks in the 1920's and 1930's do not treat the simple harmonic oscillator by operator methods but usually do so by both matrix mechanics and by wave mechanics. This includes texts like Birtwistle (1928),\cite{Birtwhistle} Condon and Morse (1929),\cite{CondonMorse29} Born and Jordan (1930),\cite{BornJordan30} Mott (1930),\cite{Mott30} Sommerfeld (1930),\cite{Sommerfeld30} Fock (1932),\cite{Fock31} Frenkel (1932),\cite{Frenkel32} Pauli (1933),\cite{Pauli33} Frenkel (1934),\cite{Frenkel34} Pauling and Wilson (1935),\cite{PaulingWilson} Jordan (1936),\cite{Jordan36} Kemble (1937)\cite{Kemble37} and Dushman (1938)\cite{Dushman}. The one exception from the 1930's appears to be Rojansky's 1938 text,\cite{Rojansky} which provides a treatment nearly identical to Dirac's 1930 method. But Rojansky makes it clear that he is working with operators (as his derivation is in a chapter entitled ``The Symbolic Method''), and he strictly works solely with the operators, never introducing the Heisenberg matrices in this section of his book (although he does discuss matrix mechanics elsewhere). While he has all of the elements available to construct the eigenfunctions abstractly in terms of the raising operators, he fails to do so. He does, however, employ the intertwining relationship in the derivation, making it closer to the way we proceeded here.

Intriguingly, Schr\"odinger\cite{schroedinger4041} developed his factorization method in 1940-41. The first problem he tackled was the simple harmonic oscillator. In this work, he showed that one can evaluate the equation $\hat a|0\rangle=0$ for the ground state (in coordinate space) and find a first-order differential equation for the ground-state wavefunction. He then simply states that one can extend the same method to higher eigenstates but provides no details. Hence, Schr\"odinger was, perhaps aptly, the first to determine all the eigenfunctions (and the associated wavefunctions) for the simple harmonic oscillator via the operator-based approach.

The next development of the operator method for the simple harmonic oscillator appears in the 1947 edition of Dirac's textbook.\cite{Dirac47}~ This gives the origin of the modern approach adopted by all subsequent textbooks and provides the complete abstract derivation. Dirac explicitly defines dimensionless operators
\begin{equation}
\eta = \sqrt{\frac{1}{2m\hbar\omega}}(\hat{p}+im\omega\hat{q})~~~\text{and}~~~
\bar{\eta} = \sqrt{\frac{1}{2m\hbar\omega}}(\hat{p}-im\omega\hat{q}),
\end{equation}
which he uses to establish this modern operator method. He checks that
\begin{equation}
\bar{\eta}\eta-\eta\bar{\eta}=1
\end{equation}
and shows that $\eta$ and $\bar{\eta}$ act as ladder operators which raise and lower the energy of the harmonic oscillator in steps of $\hbar\omega$, respectively.  Dirac demonstrates that $\eta\bar{\eta}$ is a positive semi-definite operator and uses this to show that the ground state energy of the harmonic oscillator equals $\frac{1}{2}\hbar\omega$. He expresses the $n$th energy eigenstate as $\eta^n | 0 \rangle$ and represents the wavefunctions by 
\begin{equation}
\langle q' | \eta^n | 0 \rangle
\end{equation}
which he finds using differential equations. While Dirac's method here is identical to the modern operator method used today, his notation differs slightly. He uses $\eta$ and $\bar{\eta}$ to denote the ladder operators and again includes the factor of $i$ on the position operator. It is fair to say that it is here, in 1947, that the complete abstract formulation of the simple harmonic oscillator is born.

The remainder of the harmonic oscillator's development consists mainly of notational changes. Leonard Schiff introduced, but did not significantly use, the $\hat{a}$ and $\hat{a}^{\dagger}$ notation in his 1949 quantum textbook.\cite{Schiff49}~ We suspect the reason for the use of this letter to denote the ladder operators may lie in the second volume of Sin-Itiro Tomonaga's 1953 quantum textbook.\cite{Tomonaga}` Tomonaga uses $A_s$ to denote the complex time-dependent amplitude of a De Broglie wave packet
\begin{equation}
\Psi(x,y,z,t) = \sum\limits_{s=1}^{\infty}A_s(t)\phi(x,y,z)
\end{equation}
He then gives the real and imaginary parts of $A_s$ by
\begin{eqnarray}
\Re A_s &=& \frac{1}{2}(A_s+A_s^{*}) = \sqrt{\pi}Q_s \\
\Im A_s &=& \frac{1}{2i}(A_s-A_s^{*}) = \sqrt{\pi}P_s 
\end{eqnarray}
which bears a striking resemblance to the way many popular textbooks relate $\hat{p}$ and $\hat{q}$ to the ladder operators. If our suspicions hold true, the use of $a$ would then stand for ``amplitude.'' Born and Jordan's textbook\cite{BornJordan30} also seems to support this notion, as they explicitly refer to $b$ and $b^{\dagger}$ as ``komplexe Amplituden." One should also note that Frenkel's 1934 book\cite{Frenkel34} discusses many of these same themes too when quantizing light, including the same modern notation as used by Schiff fifteen years later. Frenkel's approach is deeply entrenched in matrix mechanics, as was much of the work at that time---our interpretation is that the objects he works with are in fact matrices and not abstract operators in their full generality---but this conclusion is {\it not} crystal clear. Interestingly, Frenkel also employs the $a,~a^\dagger$ notation when quantizing light.

Through the 1950s and 1960s, we see textbooks use a combination of differential equations and Dirac's 1947 operator method to treat the harmonic oscillator. While every book's operator treatment follows Dirac's, we see a swath of different notations. We found this in Bohm,\cite{Bohm50} Landau and Lifshitz,\cite{LandauLifshitz} Messiah,\cite{Messiah} Dicke and Wittke,\cite{DickeWittke} Merzbacher,\cite{Merzbacher} Powell and Crasemann,\cite{PowellCrasemann} Harris and Loeb,\cite{HarrisLoeb} Park,\cite{Park} Gottfried,\cite{Gottfried}  Green, \cite{Green65} Ziman\cite{Ziman} and Fl\"ugge\cite{Flugge}. From the late 60s through today, all quantum textbooks use the same notation as Schiff, Messiah and Park. These include Saxon,\cite{Saxon} Baym,\cite{Baym} Gasiorowicz,\cite{Gasiorowicz} Cohen-Tannoudji\cite{Cohen-Tannoudji} and Winter\cite{Winter} in addition to virtually all subsequent textbooks. We could not figure out why all textbooks adopted a standardized notation after 1970, but the earliest instance of the modern approach with the modern notation seems to be in Messiah's 1959 textbook.\cite{Messiah}

In summary, we see the operator method for the simple harmonic oscillator to have developed as follows. The matrix mechanics approach of Heisenberg\cite{Heisenberg25} and Born and Jordan\cite{BornJordan25} already has about one third of the abstract method worked out. That approach uses the positivity of the Hamiltonian and a ladder structure of the matrix elements to determine the energy eigenvalues. The ladder operation structure was even illustrated graphically by Birtwistle.\cite{Birtwhistle}~ Next, Born and Jordan's 1930 textbook\cite{BornJordan30} was the first to represent the ladder operators in the matrix mechanics formalism, but Dirac's 1930 textbook\cite{Dirac30} initiated the abstract operator approach with the factorization of the Hamiltonian in terms of operators, even though it later employed the matrix mechanics methodology to determine the eigenvalues. Rojansky\cite{Rojansky} performed the first completely abstract derivation free from matrix mechanics. Though he was on the precipice of also determining the eigenvectors, he did not. That had to wait for Fock space\cite{Fock32} and Schr\"odinger's use of it in his factorization method\cite{schroedinger4041} before one could construct the eigenfunctions abstractly (but the derivation still required going to coordinate space to be completed). Finally, Dirac finished the modern derivation in his 1947 text.\cite{Dirac47}~ The operator method was immediately adopted by nearly all other textbooks, although the notation did not become the standard one we are accustomed to until the early 70s. Now, it has been completed with an algebraic derivation of the coordinate-space and momentum-space wavefunctions presented here.

\section{Conclusion}

The simple harmonic oscillator is generally viewed as one of the most important problems in quantum mechanics. The operator-based solution of the energy eigenvalues and eigenstates (along with the abstract methodology used to evaluate matrix elements) is often the highlight of a quantum-mechanics course. In this work, we tweaked the derivation of the eigenvalues and eigenfunctions to put them in a more standard approach motivated by the Schr\"odinger factorization methods instead of Dirac's 1947 derivation. In addition, we extended the operator-based methods to also allow for an abstract derivation of the wavefunctions in coordinate and position space. This approach employed the translation operator to shift the wavefunction from the origin and compute the change of its value. It employs simple operator identities (the Hadamard lemma and Baker-Campbell-Hausdorff identity when $[\hat A,\hat B]$ commutes with $\hat A$ and $\hat B$) and hence it is easy to understand and follow even for undergraduates in an introductory course. In addition, we explored the history behind the operator method for the simple harmonic oscillator. Our findings are that this history is much richer than simply ``didn't Dirac do that?'' Indeed, we discovered that one third of the argument can already be found in the matrix mechanics works of Heisenberg and Born and Jordan. We argue that Dirac's original 1930 treatment is much closer to the matrix mechanics approach and that it actually was Rojansky in 1938 who made the derivation a completely abstract operator argument. Even Schr\"odinger had a hand in this, being the first to use the abstract operators to construct eigenfunctions and coordinate-space wavefunctions in 1940-41. Dirac then finished the methodology in 1947. 

We hope that our completion of this work here will be adopted by others teaching quantum mechanics, as we feel it is yet another beautiful demonstration of the elegance of the abstract operator approach. Now the entire simple harmonic oscillator problem can be solved algebraically!


\begin{acknowledgments}

 Initial stages of this work were supported by the National Science Foundation under grant number PHY-1620555 and the final stages under grant number PHY-1915130. In addition, JKF was supported by the McDevitt bequest at Georgetown University. We thank Wes Mathews for a critical reading of the manuscript and Manuel Weber for  translation of sections of the original 1930 German textbook of Born and Jordan cited below. We also thank Takami Tohyama for help with determining the original Japanese reference to Tomanaga's textbook and Andrij Shvaika for help with the reference to Fock's original textbook.

\end{acknowledgments}

\end{document}